# Phishing Attacks Detection
# A Machine Learning-Based Approach

Fatima Salahdine[1,2], Zakaria El Mrabet[1], Naima Kaabouch[1]
[1]School of Electrical Engineering & Computer Sciences, University of North Dakota
Grand Forks, ND-58203, USA
[2]Department of Electrical and Computer Engineering, the University of North Carolina at Charlotte,
Charlotte, NC-28223, USA
{fatima.salahdine, zakaria.elmrabet, naima.kaabouch}@und.edu

*Abstract-* Phishing attacks are one of the most common social engineering attacks targeting users' emails to fraudulently steal confidential and sensitive information. They can be used as a part of more massive attacks launched to gain a foothold in corporate or government networks. Over the last decade, a number of anti-phishing techniques have been proposed to detect and mitigate these attacks. However, they are still inefficient and inaccurate. Thus, there is a great need for efficient and accurate detection techniques to cope with these attacks. In this paper, we proposed a phishing attack detection technique based on machine learning. We collected and analyzed more than 4000 phishing emails targeting the email service of the University of North Dakota. We modeled these attacks by selecting 10 relevant features and building a large dataset. This dataset was used to train, validate, and test the machine learning algorithms. For performance evaluation, four metrics have been used, namely probability of detection, probability of miss-detection, probability of false alarm, and accuracy. The experimental results show that better detection can be achieved using an artificial neural network.

## I. INTRODUCTION

With more than 7 billion email accounts worldwide in 2021 and over 3 million emails sent per second, email services have become an indispensable way for personal and professional transactions. However, the massive use of email services has grabbed the attention of attackers as a potential field for launching successful attacks. Compromising an email account becomes challenging or almost impossible since the email service providers offer secure E2E communication. Thus, the attackers opt for using social engineering strategies to compromise email accounts by manipulating human intelligence to obtain critical and confidential information [1].

Phishing attacks perform by sending forged emails looking legitimate from an authentic entity to a victim or a group of victims [2][3]. They aim at obtaining users' confidential data or uploading malware on their machines. For instance, the attackers send an email with a redirection link to a malicious website where the user is requested to provide some sensitive data, including bank account number or login and password. The attacker can also attach a file to the fake email to be uploaded by the victim, which can automatically trigger the execution of embedded malware.

To cope with phishing attacks and mitigate their potential risks, a number of techniques have been proposed. These techniques can be classified into four categories: rule-based, white and blacklist, heuristic, and hybrid. The rule-based approach consists of using data mining techniques to train the model based on a specific dataset with a certain number of features, then extract some phishing attacks rules. For instance, a rule-based phishing attacks approach was proposed for the banking service in which several features were selected, including IP address, SSL certificate, web address length, number of dots in URL, and blacklist keywords. In [4], the authors proposed a data mining tool called Multi-label Classifier Associative Classification in which 16 features were selected, including IP address, Long URL, URL's having @ symbol, prefix and suffix, and DNS record. In [5], a rule-based technique was described, in which 17 features were selected and different classifiers were used, namely C4.5, RIPPER, PRISM, and CBA. The results show that C4.5 outperforms the other algorithms in terms of detection rate and accuracy. Rule-based approaches are easy to implement; however, they represent some shortcomings, including a low accuracy rate.

Other techniques are based on whitelist and blacklist approaches [6][7]. In [6], a white-list-based approach was proposed in which a number of features related to the legitimate websites were recorded, such as URL, IP address, and Login User Interface. When the user visits a website that does not match any entry in this list, the requested website is classified as malicious. In [7], a blacklist-based approach was proposed in which the URL of the suspicious webpage is divided into several parts and compared to a list of phishing websites. The list of suspicious websites is gathered from several sources, including spam traps and open phishing email databases. Whitelist and blacklist approaches are inefficient in dealing with new webpages that are not included in those pre-established lists. In addition, these lists require frequent updates, which can be computationally expensive.

For the heuristic techniques, feature sets are selected, and the impact of each set in increasing the detection likelihood is investigated. The tested feature sets can range from URL, IP address to HTML DOM of the webpage. For instance, a heuristic-based technique was proposed in which 20 heuristic features were selected [8]. The results show that the URL-based and HTML-based heuristics are effective, and they outperform the blacklist-based approach. In [9], a heuristic-based approach called CANTINA+ was proposed to extract the most frequent words in the webpage and search for them on a search engine. The webpage is classified as legitimate if it appears in the first results of the research since the first reported webpage is the most visited and their likelihood of being legitimate is high. However, the attacker can access these entries and make malicious webpages appear in the first search results.

A number of hybrid detection techniques have been proposed that combine the fuzzy-logic approach with other data mining techniques [10][11][12]. In [10], a hybrid approach was proposed that has an accuracy of 98.5% with 288 features. It requires a considerable number of features, which makes its implementation complex. In [11], a hybrid approach was proposed reaching an accuracy of 86.38% with 27 features. However, it was not clear how the features were extracted. A target identification algorithm was designed to identify phishing webpages [12]. It is based on third-party services to investigate in-depth the content of the suspicious link and verify its source, which may result in more processing time.

In this paper, we investigate the efficiency of the machine learning approaches in detecting phishing emails. After understanding the research problem's requirements and analyzing the training dataset, we selected three models among others, namely support vector machine (SVM), logistic regression (LR), and artificial neural network (ANN). We explored other variations with different kernel types and different architectures. The dataset used to train and test the classifiers was from real attacks launched against the email service of the University of North Dakota.

The rest of this paper is organized as follows. Section II describes the methodology of the proposed approach. Section III discusses and compares the simulation results. Finally, a conclusion is given at the end.

## II. METHODOLOGY

### A. Features selection

Usually, a typical email is composed of a header and a body. The email header has a specific structure consisting of several information related to the sender and the receiver, including their IP addresses, the subject, and the date. Regarding the email body, it has no specific format, and it can be customized and different from one email to another. However, there are some items that can be found in any typical email, such as text, link to a website, attached files, and the email's signature. Since not the entire email content is relevant in detecting legitimate emails from malicious ones, it is important to select and extract only those specific features that are used in phishing emails. In this paper, we used ten relevant features in which eight are extracted from the email body while the rest is from the email header. These features are: sender email address, attached file extension, blacklist keywords, secure socket layer (SSL) certificate, certificate authority (CA), redirection URL, hiding links, clear IP address, website traffic, and webpage age. Individual features may not reveal the legitimacy of an email but combining several features increase the likelihood of detecting potential phishing emails.

### 1) SSL certificate

When a user is requested to enter confidential data on legitimate websites, the exchanged data between the server and the end-user is encrypted, which can be achieved through the HTTP protocol with an additional secure socket layer [13]. However, most of the phishing emails include HTTP links without any supplementary secure layer exposing the data to potential unauthorized access and loss. Thus, if an email includes a secure HTTP link, then it is legitimate; otherwise, it is malicious.

$$\text{Feature 1: if} \begin{cases} HTTPs\ link \rightarrow legitimate\ email \\ Otherwise \rightarrow suspecious\ email \end{cases}$$

### 2) Certificate authority

Not every HTTPS link can guarantee a secure connection to the server and make the sensitive data undisclosed to a third party since the SSL certificate can be delivered by an unauthentic entity or self-signed. An SSL certificate is insufficient to decide if an HTTSs link is secure. Investigating the identity of the entity that issued the certificate is crucial in verifying the email's legitimacy [14]. Thus, if an SSL certificate is not delivered by a trusted and credible authority such as GoDaddy, Comodo, and Symantec, then the email is suspicious.

$$\text{Feature 2: if} \begin{cases} Authentic\ CA \rightarrow legitimate\ SSL \\ Otherwise \rightarrow suspicious\ SSL\ certificate \end{cases}$$

### 3) Blacklist keywords

Phishing emails share in common some keywords and short phrases. These keywords have a sense of urgency, including "Click Now", "Verify Now," "Valid in 24h", and "Update Now." Including such keywords in the email, the body provides clues about the illicitness of the email. In this paper, we established a list of several suspicious keywords used by the attackers to grab the attention of the victim [15]. If the email includes one or more blacklist words, then it is malicious.

$$\text{Feature 3: if} \begin{cases} Email\ word \in \{blacklist\} \rightarrow suspicious\ email \\ otherwise \rightarrow legitimate\ email \end{cases}$$

### 4) Redirection URL

Some phishing emails include a link that implicitly redirects the user to a hidden server before reaching the requested website, such as a proxy server. This server will handle the communication between the user, the malicious, and the legitimate websites [16]. GET request of the HTTP protocol is used to verify the legitimacy of an URL.

$$\text{Feature 4: if} \begin{cases} GET\ (link_{URL}) \neq link_{URL} \rightarrow suspicious\ email \\ Otherwise \rightarrow legitimate\ email \end{cases}$$

### 5) Hiding links

An alternative way to hide the actual URL website is to use hiding links, which rely on two techniques: URL shorteners and customized HTML emails. In the former, the attacker wraps the real URL in a short one such as "goo.gl", or "j.mp". In the latter, the attacker forges an HTML email with the Cascading Style Sheets and JavaScript scripts to customize the webpage link with a personalized clicked text or image. Thus, an email is suspicious if it includes a short URL.

Feature 5: if $\begin{cases} link_{URL} \text{ is } short_{URL} \text{ or } link_{URL} \text{ is hidden inside} \\ \text{image or "clicked text"}) \rightarrow suspicious\ email \\ Otherwise \rightarrow legitimate\ email \end{cases}$

*6) Clear IP address*

Some phishing emails include links with a clear IP address. "https://50.10.125.26/index.php" is an example that indicates the illegitimacy of the email. Using an IP address instead of the specific domain name is because malicious webpage links last for less than three days, and attackers do not buy a domain name for a short period of time. Thus, if a link includes a clear IP address, then it is suspicious.

Feature 6: if $\begin{cases} link_{URL} \text{ includes IP address} \rightarrow suspicious\ email \\ Otherwise \rightarrow legitimate\ email \end{cases}$

*7) Website traffic*

Legitimate websites receive a number of requests with a specific traffic rate per day. A legitimate website has a rank less or equal to 150,000 in the Alexa database. However, phishing websites are not often visited as they have a short lifetime, and their traffic is low.

Feature 7: if $\begin{cases} traffic < 150000 \rightarrow legitimate\ email \\ Otherwise \rightarrow suspicious\ email \end{cases}$

*8) Age of the webpage*

Since most phishing webpages have a short lifetime, the age of the webpages can provide information about their legitimacy. The age of the authentic website is usually more than one year. Thus, if the email includes a webpage link with less than one year, then it is suspicious.

Feature 8: if $\begin{cases} webpage\ age > 1\ year \rightarrow legitimate\ email \\ Otherwise \rightarrow suspicious\ email \end{cases}$

*9) Sender's email address*

In some phishing emails, there is an inconsistency between the email subject and the address of the sender. For instance, some malicious emails seem to be emitted by an authentic entity, such as Microsoft or Dropbox, since the email's subject states something similar to "the user X has shared some files with you" or "Reinitialize the password." However, the sender's email address includes a strange domain name such as "@sharing.dboxfile.com" or "@dropbox.com." Thus, such inconsistency can be relevant in detecting malicious senders. Thus, if a domain name does not belong to the credible domain names list, then the email is suspicious.

Feature 9: if $\begin{cases} @domain\ name \in \{list\ of\ credible\ domaine \\ names\} \rightarrow legitimate\ email \\ Otherwise \rightarrow suspicious\ email \end{cases}$

*10) Attached file extension*

It is used to increase the likelihood of detecting phishing emails. Some phishing emails include an attached file, including an embedded payload. This payload can be an executable shell script giving the attacker the privileges to execute some command on the user's machine. One of the known tools used by attackers to forge phishing emails is the social engineering Toolkit installed by default on the Kali Linux. It generates a file including the payload with ".exe" or ".dll" extension. If the attached file has ".exe" extension, then the email is suspicious.

Feature 10: if $\begin{cases} file\_name.\text{exe} \rightarrow suspicious\ email \\ Otherwise \rightarrow legitimate\ email \end{cases}$

*B. Classification techniques*

In this work, we compared the performance of three classifiers, namely support vector machine (SVM), logistic regression (LR), and artificial neural network (ANN). SVM is a machine learning algorithm used for solving classification and regression problems [17]. It is based on a hyper line classifier that separates and maximizes the margin between two distance classes. Let the dataset, *D*, be given as {($x_1$, $y_1$), ($x_2$, $y_2$),…,( $x_N$, $y_N$)}, where $x_i$ is the set of training tuples with the associated class labeled $y_i$. Each $y_i$ can take one of two values, either +1 or -1, corresponding respectively in our case to the class 'phishing email' or 'legitimate email'. SVM finds the best decision boundary to separate these two classes using a hyper line, *h*, which can be defined as

$$h(x) = W * X + b = \sum_{i=1}^{N} \alpha_i y_i (x_i, x) + b \quad (1)$$

where *W* is the weight, *b* is the bias, *N* is the number of features in the dataset, $x_i$ is the set of training tuples, and $\alpha_i$ is the Lagrange multiplier. In the case of non-linear data, one can first transform the data through non-linear mapping to another higher dimension space and then use a linear model to separate the data. The mapping function is done by a kernel function *K* and the equation can be rewrite the equation (1) as

$$h(x) = W * X + b = \sum_{i=1}^{N} \alpha_i y_i K(x_i, x) + b \quad (2)$$

where $K(x_i, x)$ is the kernel function. In this paper, we used the polynomial function as a kernel. SVM classifies a new email based on its position with respect to that hyper line. If an email's features lie on or above the hyper line, then it belongs to the phishing email class.

LR is a supervised machine learning technique used for predicting discrete output class, classification, and binary classification [3]. It is based on different hypothesis functions for predicting a binary-value output. In this paper, sigmoid function is considered as a hypothesis function. It is given by

$$h_w(x^{(i)}) = \frac{1}{1 + e^{-\sum_{i=0}^{N} w_j x^{(i)}_j}} \quad (3)$$

where $w_j$ is the weight associated with each input $x_j$ and $N$ is the number of features. In this paper, we opted for gradient descent (GD) as an optimization technique to define the appropriate weight that minimizes the prediction error.

ANN is a supervised machine learning algorithm used for classification and regression prediction. It is composed of an input layer, one or multiple hidden layers, and an output layer where each layer is composed of several neurons. A neuron is a computation unit that takes a set of inputs associated with weights and predicts the output using an activation function. There are several activation functions, including the sigmoid function, hyperbolic tangent function, and rectified linear unit function [15]. Training an ANN model involves forward propagation and backward propagation. For each instance in the dataset, the forward propagation is used to compute the predicted output and compare it with the actual one and then calculate the error between these two values. To minimize the error, the backward propagation updates the weights associated with each input using gradient descent. Forward and backward propagation are repeated until ANN reaches a minimum error value. $i^{th}$ neuron of the $l^{th}$ layer is given by

$$a_j^{(l)} = g_j^{(l)}(\sum_{i=1}^{m} w_{ji}^{(l)} a_i^{l-1} + b_{ji}^{(l)}) \quad (4)$$

The activation function of the output layer of an ANN with one neuron is given as follows:

$$h_w(x) = g(\sum_{i=1}^{m} w_i^{(l)} a_i^l) \quad (5)$$

ANN learn their weights and biases using GD technique, Given a training set $\{(x^{(1)}, y^{(1)}), ..., (x^{(m)}, y^{(m)})\}$, the cross-entropy cost function $J(W)$ is given by

$$J(W) = -\frac{1}{m}\sum_{i}^{m}\sum_{k=1}^{K} y_k^{(i)} \log(h_w(x^{(i)})_k) + (1 - y_k^{(i)})\log(1 - h_w(x^{(i)})_k) + \frac{\lambda}{2m}\sum_{l=1}^{L-1}\sum_{i=1}^{s_l}\sum_{j=1}^{s_l+1}(w_{j,i}^{(l)})^2 \quad (6)$$

where $\lambda$ is the regularization parameter, $m$ is the training data size, $K$ is the number of the output classes, and $h_w$ is the hypothesis function, and $w_{j,i}^{(l)}$ is the weights assigned to the link between the $ith$ and $jth$ neurons of $lth$ layer.

The process consists of minimizing the cross-entropy cost function $J(W)$. Backpropagation aims at updating all the weights simultaneously to minimize the cost function. The hypothesis is the case of a sigmoid function given as:

$$(z) = \frac{1}{1+e^{-z}} \quad (7)$$

where $z$ is the vector of weights associated with the vector of features $x$. In this binary classification, there are two cases based on the values of $z$: (i) if $z \gg 0$, the hypothesis function satisfies $h_w(x) > 0.5$, which corresponds to the presence of the attack ($y = 1$); (ii) if $z \ll 0$, the hypothesis function satisfies $h_w(x) < 0.5$, which corresponds to the absence of the attack ($y = 0$).

### III. RESULTS AND DISCUSSION

To train, validate, and test the models, we built a dataset consisting of 4000 real phishing emails. These emails were collected from the North Dakota email system from May 22, 2017, to June 20, 2018. The collected data include some redundant emails because some attackers sent the same forged email to multiple users, or they used it to conduct the same attack several times. Thus, we analyzed and improved the dataset by removing the duplicated and redundant emails and reducing the number of instances to 2000 phishing emails. The legitimate emails were collected from legitimate accounts and emitted by an authentic entity. To keep the number of phishing and legitimates emails equally distributed in the dataset and to avoid bias towards any one of these two types of classes, 2000 legitimate emails were selected. Thus, the final dataset contains 4000 instances with legitimate and phishing emails, as presented in Table I.

TABLE I. COLLECTED PHISHING DATASET

| | |
|---|---|
| Total samples | 4000 |
| Total phishing emails | 2000 |
| Total legitimate emails | 2000 |
| Total training samples | 2800 |
| Total testing samples | 1200 |
| Total number of features | 10 |

Since some classifiers cannot be trained on categorical data, the dataset went through a pre-processing process in which all the nominal values were converted into numerical values. The same converting model was used to map the nominal data to the nominal one in the entire dataset. In addition, the dataset went through a feature scaling process to make the data normally distributed with zero as a mean and a standard deviation of 1. These processes can reduce the processing time for some classifiers along with avoiding the divergence issues that could arise. The performance evaluation of the algorithms was conducted using several metrics: $P_d$, $P_{fa}$, $P_{md}$, and accuracy. $P_d$ is the likelihood to detect suspicious emails when they are suspicious. $P_{fa}$ is the likelihood to detect a suspicious email while it is legitimate. $P_{md}$ is the likelihood to classify a legitimate email when this email is suspicious. The accuracy is the likelihood that a classifier attributes legitimate email to the class of "legitimate email". These metrics are expressed as

$$Pd = \frac{number\ of\ detected\ suspicious\ emails}{Number\ of\ suspicious\ emails} \quad (8)$$

$$Pfa = \frac{number\ of\ false\ detected\ suspicious\ emails}{Number\ of\ suspicious\ emails} \quad (9)$$

$$Pmd = \frac{number\ of\ miss\ detected\ susepcious\ emails}{Number\ of\ suspicious\ emails} \quad (10)$$

Examples of results are presented in Table II through Table IV. ANN performance is affected by many parameters, including the number of hidden layers, the number of hidden neurons in each layer, and activation function. To find the right set of parameters that maximize the ANN performance, we conducted several experiments using the generated dataset with different combinations of these parameters.

Examples of results are represented in Table II. As it can be seen, ANN with two hidden layers of 100 neurons, each with relu function, has the best performance as it achieves the highest $P_d$ of 90.3%, the lowest $P_{md}$ of 9.7%, and the highest accuracy of 94.5%. Thus, ANN with two hidden layers of 100 neurons each and relu activation function has the best performance.

TABLE II. ANN PERFORMANCE EVALUATION

| Algorithm | $P_d$ | $P_{fa}$ | $P_{md}$ | Accuracy |
|---|---|---|---|---|
| (100) / Relu function | 90.10% | 1.40% | 9.90% | 94.40% |
| (100,100) / Relu function | 90.30% | 1.50% | 9.70% | 94.50% |
| (100) / tanh function | 90% | 1.40% | 10% | 94.30% |
| (100,100) / tanh function | 90.10% | 1.50% | 9.90% | 94.30% |
| (100) / sigmoid | 88.90% | 1.40% | 11.10% | 93.80% |
| (100,100) / sigmoid | 88.70% | 1.40% | 11.30% | 93.70% |

TABLE III. SVM'S PERFORMANCE WITH SEVERAL KERNELS

| Algorithm | Pd | Pfa | Pmd | Accuracy |
|---|---|---|---|---|
| Linear SVM | 29.8% | 44.8% | 70.2% | 42.6% |
| Cubic SVM | 63.4% | 54.5% | 36.6% | 54.4% |
| RBF SVM | 82.3% | 27.7% | 17.7% | 77.3% |
| Sigmoid SVM | 43.3% | 24.7% | 56.7% | 59.4% |

As the performance of SVM dependents on the kernel used for email classification, four kernels were considered, namely: linear, polynomial, radial basis function (RBF), and sigmoid kernels. Examples of results are presented in Table III. Through comparing the performance of the SVM algorithms, we can conclude that SVM based RBF kernel achieves a $P_d$ of 82.3%, $P_{fa}$ of 27.7%, $P_{md}$ of 17.7%, and overall accuracy of 77.3%. Thus, it provides better results compared to the other algorithms.

To investigate the impact of the regularization parameter on the LR performance, several kernels were performed using different values of this parameter. Examples of results are given in Fig. 1 through 5. Fig. 1 represents $P_d$ against the regularization parameter. It can be seen that $P_d$ increases with the increase of the regularization parameter, reaching its maximum at 0.006 with 87.2%. For values higher than 0.06, $P_d$ decreases slightly but it remains constant at 87.1%.

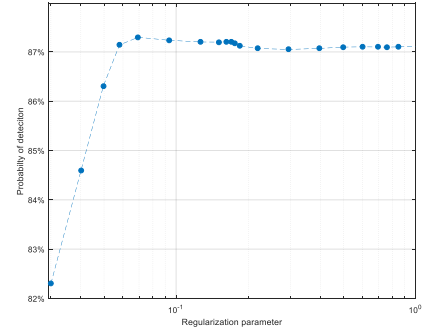

Fig.1. $P_d$ as a function of the regularization parameter.

Fig. 2 represents $P_{fa}$ as a function of the regularization parameter. As one can see $P_{fa}$ has three different regimes. For the range [0, 0.1], $P_{fa}$ is constant with an average equal to 6.5%. For the range [0.1, 0.4], $P_{fa}$ is decreasing with the increase of the regularization parameter to reach its lowest values at 0.4 with 1.4%. For values higher than 0.4, $P_{fa}$ remains constant at 1.4%.

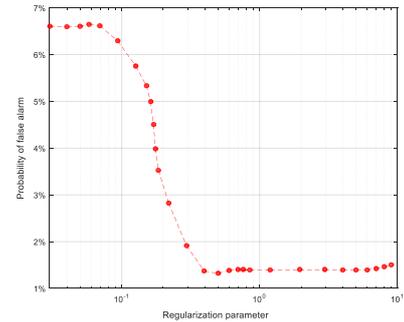

Fig.2. $P_{fa}$ versus the regularization parameter of logistic regression.

Fig.3 represents $P_{md}$ against the regularization parameter of LR. It can be seen that for the range of [0, 0.08], $P_{md}$ decreases with the increase of the regularization parameter to reach its minimum at 0.08 with 12.8%. However, for values higher than 0.08, increasing the regularization parameter does not have any impact on $P_{md}$ as it remains constant at around 12.8%.

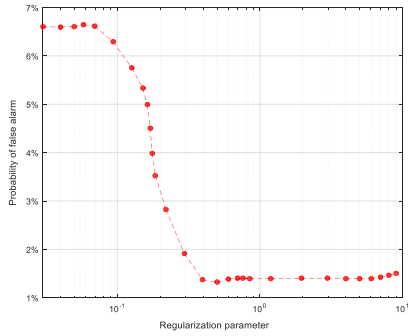

Fig.3. $P_{md}$ versus the regularization parameter of logistic regression.

Fig. 4 represents the accuracy as a function of the regularization parameter. One can see that the accuracy increases when the regularization parameter is less than 1, while it is constant for values higher than 1. It reaches its maximum value of 92.9% when the regularization parameter is 0.4. Thus, LR represents better performance with a regularization parameter higher than 0.7.

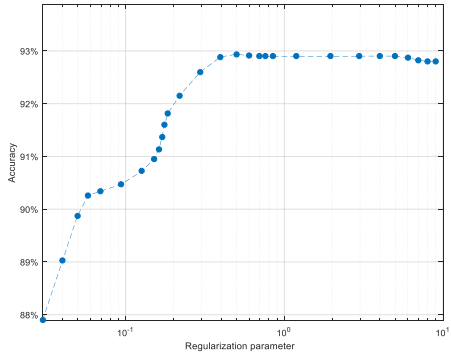

Fig.4. Accuracy versus the regularization parameter of logistic regression.

TABLE IV. COMPARISON BETWEEN ANN, SVM, AND LR

| Algorithm | Pd | Pfa | Pmd | Accuracy |
|---|---|---|---|---|
| ANN (100,100) Relu function | 90.3% | 1.5% | 9.7% | 94.5% |
| SVM Gaussian Radial basis function | 82.3% | 27.7% | 17.7% | 77.3% |
| LR regularization parameter=0.7 | 87.1% | 1.4% | 12.9% | 92.9% |

Table IV evaluates the performance of the three classifiers based on the four metrics. For ANN, we selected two hidden layers with 100 neurons, each with the Relu activation function since it produces the best results compared to other activation functions. Regarding SVM, Gaussian Radial basis kernel is selected since it produces better results in terms of $P_d$, $P_{fa}$, $P_{md}$, and accuracy. For LR, when a regularization parameter equal to 0.7, it produces the best results. Based on the best performance of each classifier, a performance comparison between these algorithms is given in Table IV. As one can see, ANN with two hidden layers with Relu function has the highest $P_d$ and accuracy, the lowest $P_{fa}$ and $P_{md}$ compared to LR and SVM.

## CONCLUSION

In this paper, we proposed a phishing attack detection technique using machine learning. Three classifiers are trained and tested on the dataset. For each classifier, a parametric study is conducted, and the best results are reported for evaluation. For SVM, high accuracy is reported by Gaussian Radial basis function kernel. For LR, the high accuracy is given by a regularization parameter corresponding to 0.4. For ANN, high accuracy is achieved with two hidden layers, 100 neurons each, and with the Relu activation function. Therefore, the proposed model allows fast and accurate phishing attacks detection.